\documentclass[aps,prl,twocolumn,10pt]{revtex4-1}
\usepackage{amsmath}
\usepackage{amsfonts}
\usepackage{amssymb}
\usepackage{graphics}
\usepackage{graphicx}

\newcommand{\cL}{{\mathcal{L}}}
\newcommand{\sgn}[1]{{\mbox{sgn} #1}}
\newcommand{\deltat}{{\tilde{\delta}}}
\newcommand{\mF}{\mathcal{F}}
\newcommand{\prs}[1]{{\left(#1\right)}}
\newcommand{\sqb}[1]{{\left[#1\right]}}
\newcommand{\prob}[1]{{\mathcal{P} \prs{#1}}}
\newcommand{\avg}[2]{{\left<#1\right>_{#2}}}
\newcommand{\vc}[1]{{\boldsymbol{#1}}}

\begin{document}

\title{Physical Laws with Average Symmetry}
\author{Alamino, R.C.}
\affiliation{Non-linearity and Complexity Research Group, Aston University, Birmingham, UK}

\begin{abstract}
This Letter probes the existence of physical laws invariant only in average when subjected to some transformation.   
The concept of a symmetry transformation is broadened to
include corruption by random noise and average symmetry is introduced by considering 
functions which are invariant only in average under these transformations.
It is then shown that actions with average symmetry
obey a
modified version of Noether's Theorem with dissipative currents. The relation of this with possible violations of physical
symmetries, as for instance Lorentz invariance in some quantum gravity theories,
is briefly commented.
\end{abstract}

\maketitle


Symmetry, the invariance of an object's features under a transformation, is a powerful and 
unifying concept. The amount of spatial symmetry of a crystal lattice characterizes many basic properties of the corresponding 
solid \cite{Ashcroft76}. The same is also valid for the order parameters of superconductors \cite{Tsuei00} and ground state 
wave-functions of topological phases \cite{Chen11}. Gauge symmetry, for instance, is the fundamental concept underlying the 
Standard Model, our most successful physical theory to date \cite{Aitchison02}. Similarly, diffeomorphism invariance, 
the independence of the physical phenomena from the coordinates of the spacetime manifold, is the greatest insight of 
general relativity. 

Even the failure of a theory's symmetry to manifest in the ground state has profound consequences. This kind of symmetry breaking 
is what generates masses in gauge bosons via the well-known Higgs mechanism \cite{Englert64,Higgs64},
a fundamental ingredient of the electroweak theory \cite{Weinberg67}.

In 1918, Noether obtained one of the most significant results concerning symmetries in physics. She proved that continuous symmetries of an action allow, under certain general conditions, the
derivation of conserved charges \cite{Noether18}. 
Since then, the theorem has been extended in several directions, giving origin for instance to the Ward-Takahashi identity in
quantum field theory \cite{Peskin95} with recent attempts to extend it to discrete transformations \cite{Capobianco11}. 

This work is concerned with a different kind of generalization and its physical consequences.
Realistically, symmetry properties of physical systems can only be assessed through experimental measurements, 
which are always subjected to some uncertainty. This uncertainty can usually  
be modeled by a stochastic process. What happens then if a symmetry is satisfied only in average? In order for this question 
to even make sense, we need to make precise the meaning of ``satisfied in average''.

In some modern quantum gravity theories, for instance, it is suggested a possible
violation of Lorentz symmetry \cite{Collins04,Sotiriou09,Blas10}, which is however not observed experimentally, among a series 
of other theoretical counter-arguments.
Still, as results obtained by putting together many
experimental measurements rely on their averaging, wouldn't it possible that what we interpret as Lorentz invariance is 
actually an invariance only of average quantities? This would open the possibility that Lorentz violating theories could, under
some assumptions, still be worth considering. If this could be true for Lorentz invariance, it could be true for
other symmetries too. What would be the consequences?

In order to answer this, we will generalize the concept of a symmetry transformation to
include a stochastic element. Transformations of a system, and therefore its symmetries, are usually treated abstractly by group 
theory. Groups of transformations can be defined by the the invariance of some properties of a system subjected to them. 
The Lorentz group is an example of transformations applied to actual spacetime coordinates, but transformations and symmetries 
can also be related to internal degrees of freedom, as in the case of gauge symmetries. 

The concept of a group, however, does not describe transformations corrupted by noise. Although in many practical cases noise
can be minimized, avoided or simply ignored in an ideal treatment, we want to analyze situations where noise is an integral 
part of the physical description and, therefore, will have to find an appropriate
generalization of the concept of group.


\noindent\emph{Average Symmetry -} When studying symmetries of physical actions, it is usual to consider only the initial and final states of the transformed
system, paying little attention to the intermediate ones. From the mathematical point of view, this is justifiable given the
properties of Lie groups \cite{Jones98}, which include most interesting transformations of physical systems. 
Physically, however, there is no reason to ascertain that this must be the case in general. 

Consider the spin-statistics theorem. The fact that exchanging particle positions needs to be considered as an actual process
of moving one particle around the other instead of just abstractly exchanging their coordinates has deep consequences. In
2D systems, we have to move from representations of the permutation group to the braid
group, leading directly to the concept of anyons \cite{Wilczek82}.

The original version of Noether's Theorem of interest to physics is proven for connected Lie groups. The properties of
continuity and connectedness are extremely important as they guarantee that, if the theorem is true for an infinitesimal 
transformation, it is also true for a finite one \cite{Olver95}. A path on the group manifold maps to a path on the 
configuration space of the system. Usually, different paths are equivalent for symmetry considerations, but that might not be
true when noise is present.

When transformations are \emph{actually} applied to a system (in the active sense), the history of how
its parameters are varied might be important. For instance, for rotations in the presence
of friction, the total energy depends on how the angle is varied and not only on the initial and final states.  

Situations in which experimental setups can be reproduced just to some extent, with a certain amount of 
noise being unavoidable, are the rule rather than the exception. Noise is a random element in experiments which, in some 
situations as quantum mechanical measurements, cannot be ignored \cite{Gaeta88,Mertens93,Gavish06}.

The usual solution is to repeat an experiment many times and try to average out the
effect of noise. Quantities whose fluctuations around the mean are inversely proportional to their size and disappear in 
the thermodynamic limit, called \emph{self-averaging quantities}, are common place in statistical physics.

It is this kind of physical situation which leads to the introduction of what we call \emph{average symmetry}. 
Consider a system subjected to transformations affected by random noise. If some feature of this system, although
being not invariant for each particular transformation, is invariant when averaged over the noise distribution, 
we say that the system possess average symmetry.

This idea can be easily visualized by considering a one-dimensional system $S$,
described by the real coordinate $x$, and a real valued function $f(x)=x^2$. The function $f$ is even, meaning that it 
is invariant under the reflection $x\rightarrow-x$ through the origin. Suppose now that each time
a reflection through the origin is accomplished, it is invariably corrupted by a random rescaling given by 
$x\rightarrow -\alpha x$, where $\alpha$ is a random variable with distribution $\prob{\alpha}$. Then we have  
\begin{equation}
  f(-\alpha x)= \alpha^2 x^2,
\end{equation}
which is, in general, different from $f(x)$ unless $\alpha^2=1$. However, if we take the average of $f$ over $\alpha$, we have
\begin{equation}
  \avg{f(-\alpha x)}{} =\avg{\alpha^2}{}x^2,
\end{equation}
and if the distribution $\prob{\alpha}$ is such that $\avg{\alpha^2}{}=1$, then the function $f$ can be said to be 
\emph{invariant on average}. 
We then say that $S$ possess an average symmetry under the \emph{noisy transformation} 
$F_\alpha x=-\alpha x$ and call this transformation an \emph{average symmetry transformation} (AST).

An important mathematical question is: As symmetries characterize groups,
is there any structure characterized by
average symmetries? It turns out that we can give a stochastic generalization of groups 
such that all group properties are recovered in the noiseless limit. 

Define a \emph{noisy transformation} as a function $F_\alpha:M\rightarrow M$, where $M$ is the configuration space of a
system $S$, depending on a random
parameter $\alpha\in\Omega$ such that, \emph{each time} the transformation is applied to an element of $M$, this parameter is drawn from a probability distribution $P_\alpha$. $F_\alpha$ is an AST if there is a function $f:M\rightarrow N$ such that
\begin{equation}
  \avg{f\prs{F_\alpha x}}{\alpha}=f(x), \quad x\in M, \alpha\in\Omega.
\end{equation}

If $G_\beta$, is another AST of $S$ with $\beta$ independent from $\alpha$, the composition $H_{\beta\alpha}=G_\beta\circ F_\alpha$ applied to $f$ gives
\begin{equation}
  \begin{split}
    \avg{f\prs{H_{\beta\alpha}x}}{\alpha,\beta} &= \avg{f\prs{G_\beta F_\alpha x}}{\alpha,\beta}\\
                                                &= \avg{\avg{f\prs{G_\beta F_\alpha x}}{\beta}}{\alpha}\\
                                                &= \avg{f\prs{F_\alpha x }}{\alpha} = f(x),
  \end{split}
\end{equation}
and is also an AST. Notice that statistical independence of $\alpha$ and $\beta$ is a
sufficient but not necessary condition for this. We now take the set
$\Gamma$ of all ASTs of $S$ and their compositions defined by the average invariance of some function or
functions and call it a \emph{noisy group}. An element    
$F_a\in\Gamma$ of this noisy group will depend on a multidimensional random parameter $a\in O=\bigcup_{n=1}^\infty \Omega^n$.  

If ASTs are applied sequentially in time, we say that a noisy group
in which each component of a multidimensional random parameter is an independent random variable is 
\emph{memoryless}. When correlations between the variables
are present, the closure of $\Gamma$ under composition becomes a more involved concept. For simplicity, we  
only consider memoryless noisy groups and drop the word ``memoryless'' for brevity. 
Associativity under composition is straightforwardly satisfied in all cases.

Because the random parameter is 
independently drawn at every application of a transformation, it might not be possible to undo a transformation 
in the memoryless case. We then generalize the identity and the inverse by requiring these to be average properties. The \emph{inverse on average} is then defined as \emph{any} transformation obeying
\begin{equation}
  \avg{I_a x }{a} = x, \qquad a\in O.
\end{equation}

It is easy to see that this definition implies
\begin{equation}
  \avg{I_a F_\alpha x}{a\alpha} = \avg{F_\alpha I_a  x}{a\alpha} =\avg{F_\alpha x}{\alpha}.
\end{equation} 

The identity is not unique. This is not unknown in generalizations like polyadic groups, where instead of a binary product we have an $n$-ary one and the identity is also not unique \cite{Post40}. The \emph{average inverse} $F^{-1}_\beta$ of $F_\alpha$ is then defined as the transformation satisfying
\begin{equation}
  \avg{F^{-1}_\beta F_\alpha x}{\alpha,\beta}=\avg{F_\alpha F^{-1}_\beta  x}{\alpha,\beta}=x,
\end{equation}
being not unique too.

It is now easy to see that, when the distributions of the random parameters become Dirac
deltas, the noiseless case, the group structure is recovered.


\noindent\emph{A Dissipation Theorem - }We now address how Noether's
Theorem is changed under average symmetries. The version we are concerned is the one stating that, in 
Hamiltonian systems whose action is invariant under  
a Lie group transformation, the Euler-Lagrange equations can be written as a gradient, resulting in
the conservation of a current that can be constructively obtained from the theorem itself. 

We will now derive
the consequences of an action which is invariant in average, or equivalently, under the action of a noisy group.
Given that noise and dissipation are closely related \cite{Callen51}, we can expect that conservation will be
compromised to some extent. We will see how and under which conditions this expectation is
fulfilled.

Noether's Theorem relies on the fact that all
elements of a Lie group are continuously connected to the identity, what makes sufficient to prove the theorem for
infinitesimal transformations. If the transformation is applied to a set of coordinates
$x$, then the new coordinates $x'$ are also continuously connected to $x$. Physically, we can imagined that by varying
continuously the parameters $\vc{\theta}$ of the Lie group transformation we create a continuous path in the group manifold
connecting the identity to $\vc{\theta}$ which is mapped to another continuous path in the configuration manifold from
$x$ to $x'(\vc{\theta})$.
 
The noisy group structure we defined is still too general. We need to restrict it in such a way that it becomes sufficient to prove the theorem only in the infinitesimal case too. One of the
requirements is that $x$ should
remain continuously connected to $x'$ during the transformation. The transformation can then be seen as a stochastic
process. The simplest stochastic process is a diffusion process and this is the kind of transformation we will consider to
affect $x$. This will guarantee that, although the path in configuration space is everywhere non-differentiable, it is
still continuous.

Consider that our configuration space is a $d+1$ dimensional spacetime 
with coordinates $x^\mu$, $\mu=0,1,...,d$. For infinitesimal transformations, the infinitesimal
change in $x$ is linear in the infinitesimal parameters of the Lie group (summation over repeated indices is assumed in the
rest of this paper unless otherwise stated)
\begin{equation}
  dx^\mu = \alpha^\mu_i(x) \, d\theta^i,
\end{equation} 

If the path is continuous 
in the group manifold, we can
parameterize it in terms of a monotonically increasing parameter $\xi$ (arc-length, for instance) and write
\begin{equation}
  dx^\mu = \alpha^\mu_i(x) \, \theta'^i(\xi) \, d\xi.
\end{equation}

The generalization to a diffusion process is then immediate and gives the Ito stochastic differential equation \cite{Gardiner04}
\begin{equation}
  dx^\mu = A^\mu(x,\xi) \, d\xi + \sqrt{D}\,dW^\mu(\xi),
\end{equation}
where $A^\mu(x,\xi)=\alpha^\mu_i(x) \theta'^i(\xi)$, $D$ is a diffusion constant and the $W^\mu$'s are
independent Wiener processes.

The Wiener processes are Gaussian distributed with zero mean and variance $\sigma^2=D\,d\xi$ for an infinitesimal transformation, 
which implies that all terms in the forthcoming expansions can be written solely in terms of $\sigma^2$ as the moments of the
Gaussian will only depend on it.

The crucial point that justifies the sufficiency of proving our extension of Noether's Theorem for infinitesimal
noisy transformations lies on the 
geometric properties of the Wiener process. Being self-similar, all statistical properties of a Wiener process are
scale invariant. This means that they are the same no matter how far we zoom in or out of the resulting path.
Therefore, every statistical property derived for an infinitesimal interval, has to be valid also for finite ones.   

We consider now a field theory with a Lagrangian density $\cL(x)\equiv\cL\prs{\phi(x),\partial_\mu\phi(x)}$, where $\phi(x)$
represents $n$ fields $\phi_r$, $r=1,...,n$ at the spacetime point $x$. We say that it has average symmetry if the average value of its action subjected to a noisy transformation $F_a^\theta$ is invariant. The vector $\theta=(\theta^1,...,\theta^M)$
represents the (non-random) parameters of the noiseless transformation and $a=(a^1,...,a^N)$ the noisy variables, with $M$ and $N$ integers.
When $a=0$, all $F_0^\theta$ form a group, which we consider to be a Lie group with $F_0^0$ its 
identity. 

We now consider an infinitesimal noisy transformation applied to the action. The presented calculations follow a similar sequence as in \cite{Greiner96} with the appropriate modifications for the stochastic case.
For convenience of notation in the expansions, let us define the full-parameter vector
$\lambda=\prs{\lambda^0,\lambda^1,...,\lambda^N}=(\theta,a)$ and define $\mF(\lambda,x)\equiv F_a^\theta x$.
The action of the noisy group on the coordinates becomes
\begin{equation}
  x'^\mu=\mF^\mu(\lambda,x), \qquad \mF^\mu(0,x)=x^\mu.
\end{equation}

We want to calculate the following variation
\begin{equation}
  \begin{split}
    \avg{\delta I}{} &= \avg{\int dx'\cL'(x')-\int dx\cL(x)}{}\\
                                 &= \int dx \avg{\cL'(x')\left|\frac{\partial x'}{\partial x}\right|-\cL(x)}{},
  \end{split}  
\end{equation}
where $\cL'(x')=\cL(\phi'(x'),\partial'_\mu\phi'(x'))$ and $|\partial x'/\partial x|$ is the Jacobian of the
transformation.

Expanding around $\lambda=0$ we obtain
\begin{equation}
  x'^\mu = x^\mu + \Delta x^\mu + \frac12 \Delta^2x^\mu + O(\lambda^3),
\end{equation}
with
\begin{equation}
  \Delta x^\mu  = \lambda^i\partial_i\mF^\mu, \qquad \Delta^2x^\mu = \lambda^i\lambda^j\partial_i\partial_j\mF^\mu,
\end{equation}
where partial derivatives are relative to the components of $\lambda$ and $O(\lambda^3)$ indicates higher order 
terms that can be ignored. Accordingly, the Jacobian matrix becomes
\begin{equation}
  J^\mu_\nu\equiv\frac{\partial x'^\mu}{\partial x^\nu} = \delta^\mu_\nu+\partial_\nu\Delta x^\mu+\frac12\partial_\nu\Delta^2x^\mu,
\end{equation}

Its determinant is
\begin{equation}
  J = \sum_{\Sigma} \sgn{\Sigma} \, \prod_{\mu} J^\mu_{\Sigma_\mu},
\end{equation}
where $\Sigma$ is a permutation of the spacetime indices.

Up to second order, the only terms that survive are those containing either all diagonal terms or $d-1$ diagonal and two off-diagonal terms. The first kind is just the product of the diagonal entries, the second is composed of
permutations that exchange two indices, which are always odd. The resulting products lead to
\begin{equation}
  \begin{split}
	  J &= 1+\partial_\mu \Delta x^\mu +\frac12 \partial_\mu \Delta^2x^\mu\\
      &  \quad+\frac12 \sqb{(\partial_\mu\Delta x^\mu)(\partial_\nu\Delta x^\nu)
         -(\partial_\mu\Delta x^\nu)(\partial_\nu\Delta x^\mu)}.
  \end{split}    
\end{equation}

We then define the two differences
\begin{align}
  \delta  f(x) &= f'(x')-f(x),\\
  \deltat f(x) &= f'(x)-f(x),
\end{align}
with $\delta f$ measuring the total variation of the function, including both its functional change and the change in the coordinates, and $\deltat f$ focusing \emph{only} on the functional change. They are related by
\begin{equation}
  \deltat f(x)=\delta f(x)-\sqb{f'(x')-f'(x)},
\end{equation}

An expansion to second order gives
\begin{equation}
  \begin{split}
    \Delta f(x) &\equiv f'(x')-f'(x) \\
                &= \sqb{\partial_\mu f'(x)}\Delta x^\mu + \frac12\sqb{\partial_\mu f'(x)}\Delta^2 x^\mu\\
                &  \quad+\frac12\sqb{\partial_\mu\partial_\nu f'(x)}\Delta x^\mu \Delta x^\nu.  
  \end{split}
\end{equation}
  
We can now write
\begin{equation}
  \begin{split}
    \avg{\delta I}{} &= \int dx\avg{\deltat\cL(x) J}{}\\
                     &  \quad+ \int dx\avg{\Delta\cL(x) J+\cL(x)\prs{J-1}}{},
  \end{split}
\end{equation}
where, using the notation
\begin{equation}
  \cL^{r\mu}\equiv \frac{\partial\cL}{\partial\phi_{r,\mu}}, \qquad 
  \cL^{r\mu s\nu}\equiv\frac{\partial^2\cL}{\partial\phi_{r,\mu}\partial\phi_{s,\nu}},
\end{equation}
we have
\begin{equation}
  \begin{split}
    \deltat\cL &= \partial_\mu\prs{\cL^{r\mu}\,\deltat\phi_r }
                  +\frac12\partial_\mu\left[\partial_\nu\prs{\cL^{r\mu s\nu}\deltat\phi_r\deltat\phi_s}\right.\\
               &  \quad\left.   
                  -\cL^{r\mu s\nu}\prs{\partial_\nu\deltat\phi_r}\deltat\phi_s\right]
                  -\frac12 \deltat\phi_r\,\partial_\nu\prs{\cL^{r\mu s\nu}\partial_\mu\deltat\phi_s}.
  \end{split}
  \label{equation:Lt}
\end{equation}

Collecting the second order terms, we have
\begin{equation}
  \avg{\delta I}{} = \int dx\avg{\Omega}{}, \qquad \avg{\Omega}{} = \partial_\mu j^\mu + \avg{\Lambda}{},
\end{equation}
where we write the current as a summation of two terms
\begin{align}
  j^\mu   &= \avg{n^\mu}{}+\avg{j_S^\mu}{},\\
  n^\mu   &= \cL^{r\mu}\,\deltat\phi_r+\cL\Delta x^\mu,\\
  j_S^\mu &= \frac12 \cL\Delta^2x^\mu+\partial_\nu\prs{\cL^{r\nu} \deltat\phi_r}\Delta x^\mu
             +\frac12 \prs{\partial_\nu\cL}\Delta x^\nu \Delta x^\mu\nonumber\\
          &  \quad 
             +\frac12 \cL\prs{\Delta x^\mu \partial_\nu\Delta x^\nu-\Delta x^\nu\partial_\nu\Delta x^\mu},
\end{align}
and the \emph{dissipative term} is
\begin{equation}
  \begin{split}
    \Lambda &=  \frac12 \left[\deltat\phi_r\deltat\phi_s\partial_\mu\partial_\nu\cL^{r\mu s\nu}+
                2\deltat\phi_r\prs{\partial_\nu\deltat\phi_s}\partial_\mu\cL^{r\mu s\nu}\right.\\
            &   \quad\left.+
                \prs{\partial_\mu\deltat\phi_r}\prs{\partial_\nu\deltat\phi_s}\cL^{r\mu s\nu}\right],
  \end{split}                     
\end{equation}
which cannot be written as a gradient. The first term in $j^\mu$ is just the averaged value of the usual
Noether current, while the second is a stochastic contribution.

Therefore, if the integrand vanishes, we have a \emph{dissipation law} given by
\begin{equation}
  \partial_\mu j^\mu=-\avg{\Lambda}{}.
\end{equation}

In the noiseless limit, when $D=0$, we fully recover the deterministic version of Noether's Theorem.


\noindent\emph{Scalar Field - }Consider a single scalar field which is invariant under noisy infinitesimal translations
\begin{equation}
  \phi'(x')=\phi(x), \qquad x'^\mu = x^\mu + \theta^\mu + a^\mu, 
\end{equation}
with $\theta^\mu$ infinitesimal constants and $a^\mu$ a Wiener process. The difference in the Lagrangian will be due exclusively to the change in the coordinates. For this case we have the
following simplifications
\begin{equation}
\Delta x^\mu = \theta^\mu + a^\mu, \qquad \Delta^2x=0.
\end{equation}
and also
\begin{equation}
\delta\phi(x)=0 \Rightarrow \deltat\phi=-\Delta x^\mu \partial_\mu\phi
                            +\frac12 \Delta x^\mu \Delta x^\nu \partial_\mu\partial_\nu\phi.
\end{equation}

Because the random parameters will only affect the coordinates, averages will only affect their variations
\begin{equation}
  \avg{\Delta x^\mu}{}=\theta^\mu, \qquad \avg{\Delta x^\mu\Delta x^\nu}{}=\sigma^2 \delta^{\mu\nu}.
\end{equation}

The average Noether current then becomes
\begin{equation}
  \avg{n^\mu}{} = \cL\theta^\mu-\cL^\mu(\partial_\nu\phi)\theta^\nu\\
                  +\frac12\sigma^2\delta^{\lambda\nu}\cL^\mu\partial_\lambda\partial_\nu\phi,            
\end{equation}
while the stochastic current gives  
\begin{equation}
  \avg{j^\mu_S}{} = \frac12\sigma^2\delta^{\mu\nu}\sqb{\partial_\nu\cL-2\partial_\lambda
                    (\cL^\lambda\partial_\nu\phi)},            
\end{equation}

The dissipative term is
\begin{equation}
  \begin{split}
    \avg{\Lambda}{} &= \frac12 \sigma^2 \delta^{\lambda\rho}\left[
    									 \prs{\partial_\lambda\phi}\prs{\partial^\rho\phi}\partial_\mu\partial_\nu\cL^{\mu\nu}\right.\\
    							  &  \quad\left.+2\prs{\partial_\lambda\phi}\prs{\partial_\nu\partial^\rho\phi}\partial_\mu\cL^{\mu\nu}
                       +\prs{\partial_\mu\partial_\lambda\phi}\prs{\partial_\nu\partial^\rho\phi}\cL^{\mu\nu}\right].
  \end{split}                    
\end{equation}

The dissipation becomes proportional to $D$ and depends on terms with up to four spacetime derivatives. For small values of
noise, the dissipative terms become very difficult to detect in regions where the field is slowly varying.
On the other hand, this suggests that this effect might be sought in strongly varying fields. 

Another difficulty in detecting effects like this would also appear if, for instance, $\Lambda=\avg{j^\mu_S}{}=0$ but
the variation of the fields still involve terms depending on $D$. In this case, an average Noether current would be
conserved. It would even be possible that that contributions from the dissipation would completely average away
implying that the average character
of the symmetry might not be observed by looking to the conservation law at all. 

The exact value of $D$ for each situation is an important question. In applications to quantum theories, it 
is probably related to quantum fluctuations. On the other hand, a different and very interesting possibility would be the
existence of a fundamental random quantum field in the space which could induce noise by means of a coupling to 
the Standard Model fields. 

There are many questions raised by the methods and results presented here. One immediate route of research to follow 
would be to extend the framework we developed for quantum instead of classical field theories. Another would be to analyze 
what kind of Lagrangians, if any, can reproduce those present in the Standard Model in
average, while not exactly. This would allow us to explore possible unnoticed violations in the symmetry laws that are usually assumed to be exact in nature.


I would like to thank Dr J.P. Neirotti and Dr L. Rebollo-Neira for very useful discussions.

\bibliographystyle{nature}
\bibliography{avs}

\end{document}